\documentclass[preprint,numerical]{revtex4}
\usepackage{graphicx, epsfig}

\def\jnl@style{\it}
\def\aaref@jnl#1{{\jnl@style#1}}
\def\aaref@jnl#1{{\jnl@style#1}}

\def\aj{\aaref@jnl{AJ}}                   % Astronomical Journal
\def\apj{\aaref@jnl{ApJ}}                 % Astrophysical Journal
\def\apjl{\aaref@jnl{ApJ}}                % Astrophysical Journal, Letters
\def\apjs{\aaref@jnl{ApJS}}               % Astrophysical Journal, Supplement
\def\apss{\aaref@jnl{Ap\&SS}}             % Astrophysics and Space Science
\def\aap{\aaref@jnl{A\&A}}                % Astronomy and Astrophysics
\def\aapr{\aaref@jnl{A\&A~Rev.}}          % Astronomy and Astrophysics Reviews
\def\aaps{\aaref@jnl{A\&AS}}              % Astronomy and Astrophysics, Supplement
\def\mnras{\aaref@jnl{MNRAS}}             % Monthly Notices of the RAS
\def\prd{\aaref@jnl{Phys.~Rev.~D}}        % Physical Review D
\def\prl{\aaref@jnl{Phys.~Rev.~Lett.}}    % Physical Review Letters
\def\qjras{\aaref@jnl{QJRAS}}             % Quarterly Journal of the RAS
\def\skytel{\aaref@jnl{S\&T}}             % Sky and Telescope
\def\ssr{\aaref@jnl{Space~Sci.~Rev.}}     % Space Science Reviews
\def\zap{\aaref@jnl{ZAp}}                 % Zeitschrift fuer Astrophysik
\def\nat{\aaref@jnl{Nature}}              % Nature
\def\aplett{\aaref@jnl{Astrophys.~Lett.}} % Astrophysics Letters
\def\apspr{\aaref@jnl{Astrophys.~Space~Phys.~Res.}} % Astrophysics Space Physics Research
\def\physrep{\aaref@jnl{Phys.~Rep.}}      % Physics Reports
\def\physscr{\aaref@jnl{Phys.~Scr}}       % Physica Scripta

\begin{document}

\title{Gravitational waves from neutron stars: Promises and challenges}

\author{N. Andersson$^1$, V. Ferrari$^2$, D.I. Jones$^1$, K.D. Kokkotas$^3$,
B. Krishnan$^4$, J. Read$^4$, L. Rezzolla$^{4,5}$ \& B. Zink$^3$}

\address{$^1$School of Mathematics, University of Southampton, Southampton SO17 1BJ, UK\\
$^2$Dipartimento di Fisica ``G. Marconi'', Sapienza Universit\` a di Roma
and Sezione INFN  ROMA1, piazzale
Aldo Moro 2, I-00185 Rome, Italy\\
$^3$Theoretical Astrophysics, Eberhard-Karls University of T\"ubingen, 
T\"ubingen 72076, Germany\\
$^4$ Max-Planck-Institut f\"ur Gravitationsphysik,
  Albert-Einstein-Institut,
  Potsdam-Golm, Germany\\
$^5$ Department of Physics and Astronomy,
  Louisiana State University,
  Baton Rouge, LA, USA}

\def\be{\begin{equation}}
\def\ee{\end{equation}}
\def\bea{\begin{eqnarray}}
\def\eea{\end{eqnarray}}
\def\f{{\bf f}-}
\def\g{{\bf g}-}
\def\gappreq{\!\stackrel{\scriptscriptstyle >}{\scriptscriptstyle \sim}\!}
\newcommand{\ms}{{\rm ms}}
\newcommand{\km}{{\rm km}}
\newcommand{\hz}{{\rm Hz}}
\newcommand{\khz}{{\rm kHz}}
\newcommand{\mpc}{{\rm Mpc}}
\newcommand{\cf}{\textit{cf.}}
\newcommand{\ie}{\textit{i.e.}~}
\newcommand{\eg}{\textit{e.g.}~}

\begin{abstract}
We discuss  different ways that neutron stars can generate
gravitational waves, describe recent improvements in modelling the relevant scenarios
in the context of improving detector sensitivity, and show how observations are
beginning to test our understanding of fundamental physics. 
The main purpose of the discussion is to establish promising science 
goals for third-generation ground-based detectors, like the Einstein Telescope, 
and identify  the various challenges that need to be met if we want to use gravitational-wave
data to probe neutron star physics.
\end{abstract}

\maketitle

\section{Context}

Neutron stars are cosmic laboratories of exotic and exciting physics.
With a mass of more than that of the Sun compressed inside a radius of about
10 kilometers, their density reaches beyond nuclear saturation. In essence,
our understanding of these extreme circumstances requires physics that cannot
be tested in terrestrial laboratories. Instead, we must try to use astrophysical
observations to constrain our various theoretical models.

We already have a wealth of data from radio, X-ray and gamma-ray observations, providing
evidence of an incredibly rich phenomenology. We have learned that neutron stars
appear in many different disguises, ranging from radio pulsars
and magnetars to accreting millisecond pulsars, radio transients and intermittent
pulsars. Our models for these systems remain rather basic, despite four decades of effort
going into modelling the violent dynamics of supernovae and gamma-ray bursts, trying to understand the
radio pulsar emission mechanism, glitches, accreting systems etcetera.
In the next few years we expect ``gravitational-wave astronomy''
to (finally) become reality. This is an exciting prospect, because gravitational-wave (GW) observations
have the potential to probe several aspects of neutron star physics \cite{2009astro2010S.229O}.
Moreover, the information gleaned will be complementary to
electromagnetic observations. In particular, we would hope to be
able to provide constraints on the state of matter at extreme densities.

In the last few years the first generation of large-scale interferometric GW detectors
(LIGO, GEO600 and Virgo) have reached the original design sensitivity in a broad frequency window \cite{abbott}. Achieving a
sensitivity to detect a  (dimensionless) strain amplitude of $h <
10^{-21}$  at a frequency  around 100 Hz is
obviously an enormously impressive feat of technology. Moreover, the detectors are very stable.
A full years worth of (triple coincidence) high quality data was taken during the LIGO S5 run. This data is
now being analysed and, even though there has not yet been a detection, the experiment has 
provided interesting information. The LIGO detectors are now running in an enhanced configuration.
In the next five year period, they will be upgraded using advanced technology. 
Once this upgrade is complete, around 2015, the second generation of ground-based detectors will
reach
the level of sensitivity where the first detection can be expected (about one order of magnitude more sensitive than the first generation). Meanwhile, the  discussion
of third generation (3G) detectors has begun in earnest with the EU funded Einstein Telescope (ET) design study.
The aim of 3G detectors is to improve the broadband sensitivity by (roughly) another order of magnitude.

The main aim of this article is to give a brief overview of how future observational capabilities
may impact on our understanding of neutron stars. We will try to identify the promising scenarios, and
how  we need to improve our theoretical models if we want to extract maximum neutron
star science from an instrument like ET.

Neutron stars radiate gravitationally in a number of ways. The most promising scenarios involve:

\noindent
\textbf{Inspiralling binaries}: The slow orbital evolution is well modelled within the post-Newtonian
approximation, and a detection would allow the extraction of the individual masses, spins etcetera. The internal
composition of the bodies will become important at some point before merger, but it is still not clear to
what extent the compressibility, or tidal resonances, lead to observable features in the GW signal.
For second generation detectors this issue may not be so important because the late stages of binary evolution
will be difficult to detect anyway. The situation will be very different for 3G detectors, for which a
key science target will be to extract as much physics information from these systems as possible.

\noindent
\textbf{Supernova core collapse}: The violent dynamics associated with a supernova core collapse is expected
to lead to GW emission through a number of channels \cite{2003LRR.....6....2N,2009CQGra..26f3001O}. The large scale neutrino
asymmetries associated with the standing accretion shock instability (SASI) should be relevant, the global dynamics
of the collapsing core through the shock bounce, and the oscillations of the hot newly born compact object
will also affect the signature. Current simulations suggest that core collapse events in our galaxy should be detectable
with current technology. Simple scaling out to distances visible with ET
suggests that, even though the event rate may still not be overwhelmingly impressive, the detection
of GWs from supernovae provides a key target. In particular,
it appears that different suggested supernova explosion mechanisms may lead to rather different
GW signals.

\noindent
\textbf{Rotating deformed stars}: Rotating neutron stars will radiate
gravitationally due to asymmetries.  The required deformation can be
due to strain built up in the crust, or indeed the deep core if it has
elastic properties, the magnetic field or arise as a result of
accretion. Our current understanding of this problem is mainly based
on attempts to establish how large a deformation the star can sustain,
e.g. before the crust breaks.  The best estimates suggest that the
crust is rather strong \cite{2009PhRvL.102s1102H}, and that it could, in principle, sustain
asymmetries as large as one part in $10^5$
\cite{2006MNRAS.373.1423H}.  However, it is important to understand
that this does not in any way suggest that neutron stars will have
deformations of this magnitude.  The real problem is to provide a
reasonable scenario that leads to the development of sizeable
deformations. In this sense, accreting systems are promising because
of the expected asymmetry of the accretion flow near the star's
surface. Of course, accreting systems are quite messy so the required
modelling is very hard.

\noindent
\textbf{Oscillations and instabilities}: Neutron stars have rich oscillation spectra which, if detected,
could allow us to probe the internal composition. The basic strategy for such ``gravitational-wave
asteroseismology'' has been set out \cite{1998MNRAS.299.1059A}, but our models need to be made much more realistic if the method
is to be used in practice. It is also important to establish why various oscillation modes
would be excited in the first place, and obviously what level of excitation one would expect.
In order to address this problem we need to be able to model potentially relevant scenarios
like magnetar flares and pulsar glitches. It seems inevitable that such events will emit GWs at some level, but at the present time we do not have any realistic estimates 
(although see \cite{2009arXiv0910.3918S}). We need more work on these
problems, even if the end result is that these systems are not promising sources. The most promising
scenarios may be associated with unstable modes. There are a number of interesting
instabilities, like the GW driven instability of the f- and r-modes, the
dynamical bar-mode and low $T/W$ instabilities, instabilities associated with a relative
flow in a superfluid core etcetera. In recent years our understanding of these instabilities has improved
considerably, but we are still quite far away from being able to make quantitative predictions.

As we will discuss, modelling these different scenarios is far from easy. Basically, our understanding
of neutron stars relies on much poorly known physics.
In order to make progress, we must
combine supranuclear physics (the elusive equation of state) with magnetohydrodynamics,
the crust elasticity, a description of
superfluids/superconductors (which is relevant since these systems are ``cold'' on the nuclear physics temperature scale) and potentially exotic phases of matter like a
deconfined quark-gluon plasma or hyperonic matter. Moreover, in order to be quantitatively accurate, all models
have to account for relativistic gravity.

It is probably  unrealistic to expect that we will be able to resolve all the involved issues 
in the next decade. Most likely, we will need better observational data to place constraints on 
the many theoretical possibilities. Thus, it is very important to consider what we can hope to learn
about the different emission mechanisms from future GW observations. This discussion is particularly 
relevant for ET (and other 3G detectors) where key design decisions still have to be made. In this context it is relevant to ask 
what we can hope to achieve with ET, but not (necessarily) with the second generation of detectors. 
How much better can you do with (roughly) a one order of magnitude improvement in broadband sensitivity?
Are there situations where this improvement is needed to see the signals in the first place, 
or is it more a matter of doing better astrophysics by getting better statistics and an increased 
signal-to-noise to faciliate parameter extraction? 

Note: various detector noise cuves are used in this article.  The
Advanced LIGO curves are taken from the LIGO technical document
LIGO-T0900288-v2    ``Advanced LIGO anticipated sensitivity curves''.
The Virgo curve is taken from \cite{2009arXiv0907.0462R}.  The
ET curve was obtanied from the website of the Working
Group 4 of the Einstein Telescope project (\texttt{https://workarea.et-gw.eu/et/WG4-Astrophysics/}). Note also that this review is not exhaustive in any sense. We have focussed on some of the (in our opinion) key aspects. We are often citing the most recent work rather than providing a complete history.

\section{Binary inspiral and merger}

The late stage of inspiral of a binary system provides an excellent GW source \cite{sathya}.
As the binary orbit shrinks due to the energy lost to radiation, the GW amplitude rises and
the frequency increases as well. This inspiral ``chirp'' is advantageous for the observer in many ways.
First of all, it is well modelled by post-Newtonian methods and does not depend (much) on the
actual physics of the compact objects involved. In fact, much of the signal is adequately described by a point-mass approximation.
From an observed binary signal one would expect to be able to infer the individual masses,
the spin rates of the objects and the distance to the source.

A key fact that makes binary systems such attractive sources is that the amplitude of the signal
is ``calibrated'' by the two masses. The only uncertainty concerns the event rate for inspirals in a given
volume of space. Given this, it is natural to discuss the detectability of these systems in terms of the
``horizon distance'' $d_h$, the distance at which a given binary signal would be observable with a given detector.
Let us assume that detection requires a signal-to-noise ratio (SNR) of 8, and focus on equal mass neutron star
binaries (we take each star to have mass $1.4 M_\odot$). For such systems, the current and predicted future horizon
distance and the expected event rates are given in Table~\ref{tab1}
below \cite{2008ApJ...675.1459K, isenberg-2008}.

\begin{table}[h]
\begin{tabular}{lll}
\hline
detector  & $d_h$ & event rate \\
\hline
LIGO S5 & 30 Mpc & 1 event per 25-400 yrs \\
Advanced LIGO/Virgo & 300 Mpc &
Several to hundreds of events per year \\
ET & 3 Gpc &
Tens to thousands of events per year\\
\hline
\end{tabular}
\caption{Horizon distances and estimated event rates for
   different generations of GW detectors.  \label{tab1}}
\end{table}

From this data we learn a number of things. First of all, we see why it would be quite surprising
if a binary neutron star signal were to  be found in the LIGO S5 data. Given even the most optimistic estimated event rate from
population synthesis models, these events would be rare in the observable volume of space.
The situation changes considerably with Advanced LIGO/Virgo. Based on our current understanding, one would expect
neutron star binaries to be seen once the detectors reach this level of sensitivity. However, it is also clear
that if the most pessimistic rate estimates are correct, then we will not be able to gather a statistically significant sample of
signals. Most likely, we will need detectors like ET to study populations.

\begin{figure*}[t]
 \begin{center}
  \vskip 1.0cm
  \includegraphics[angle=-0,width=0.475\textwidth]{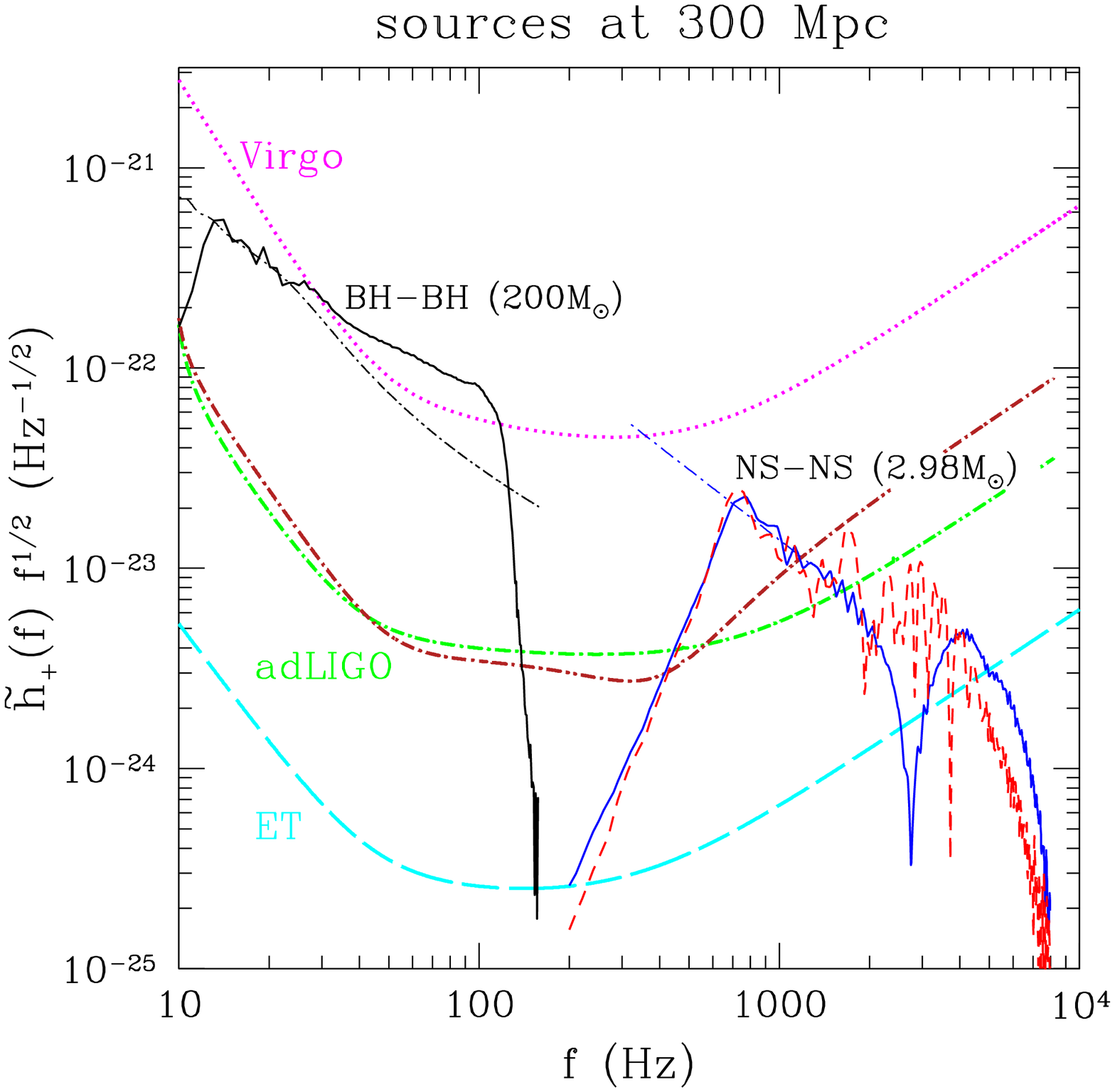}
  \hskip 0.5cm
  \includegraphics[angle=-0,width=0.475\textwidth]{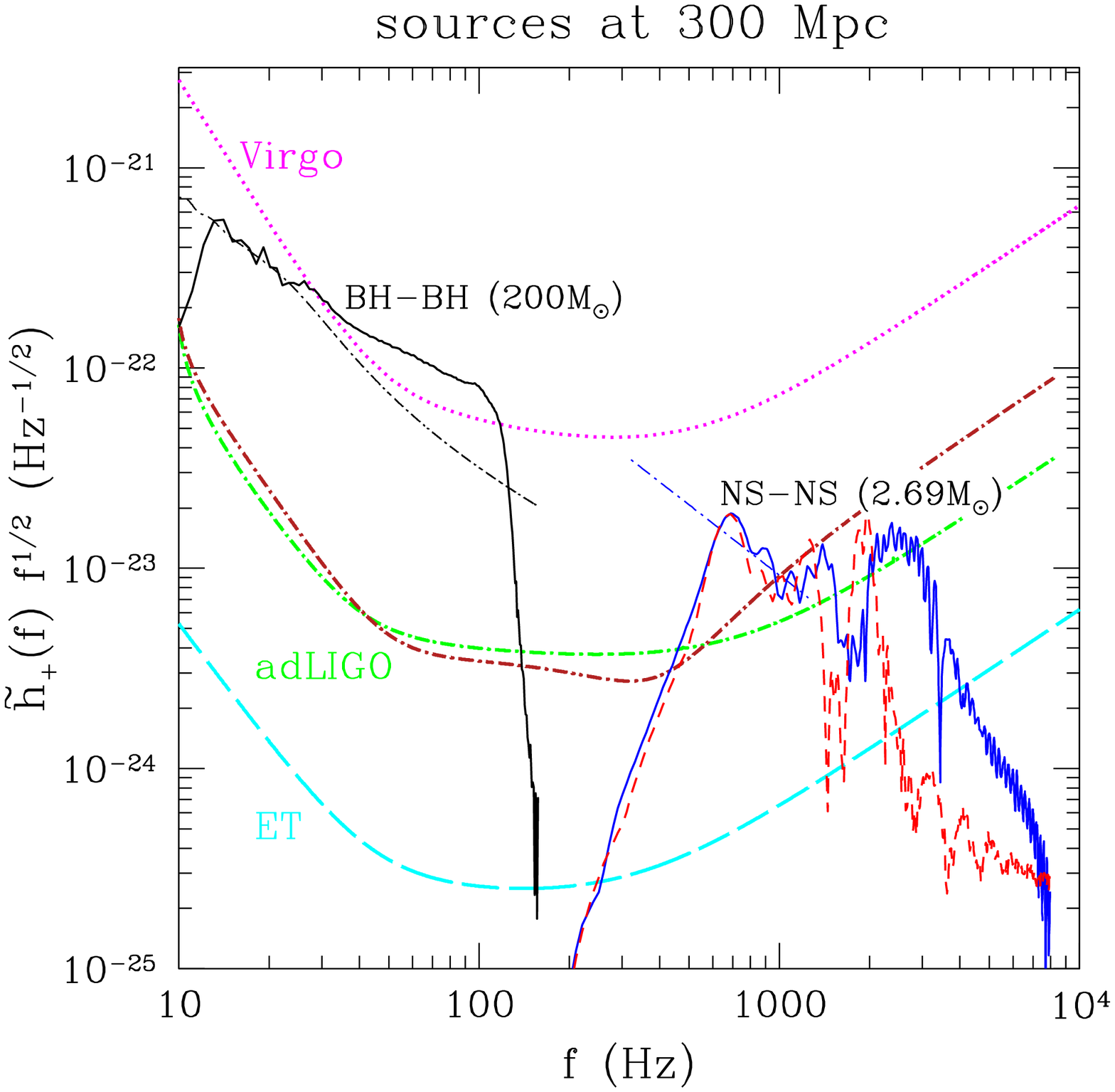}
 \end{center}
 \vskip -0.5cm
   \caption{\textit{Left panel}: PSD of the $l=m=2$ component of
     ${\tilde h}(f) f^{1/2}$ for the high-mass binary when evolved
     with the ``cold'' (blue solid line) or with the ``hot'' (red
     dashed line) EOS. Shown for comparison is  the
     corresponding PSD for an equal-mass, nonspinning black-hole
     binary with total mass $M=200\,M_{\odot}$, which appears in the
     low-frequency part of the spectrum (black solid
     line). \textit{Right panel}: The same as in the left panel but
     for the low-mass binary. In both panels we indicate (as 
     dot-dashed lines)  phenomenological evolutionary tracks
     obtained by combining PN and numerical relativity
     results~\cite{Aetal07,Aetal08}. The results are compared to the sensitivity curves of present and
     future detectors; Virgo (dotted magenta line), 
advanced LIGO (zero-detuned configuration as a green dot-dashed line and a configuration optimized for binary neutron star inspirals as a brown dot-dashed
line) and ET (as a long-dashed, cyan line).}
   \label{fig:hs_psd_pol_vs_IF}
\end{figure*}

A third generation of detectors is also likely to be required if we want to study the final stages of inspiral, including the
merger. This is a very interesting phase of the evolution, given that
the merger will lead to the formation of a hot compact remnant with violent dynamics. It may also trigger a
gamma-ray burst \cite{2003ApJ...585L..89K,2003ApJ...589..861K,2006ApJ...641L..93F,2009ApJ...702.1171C,2009CQGra..26t4016C}. Most of this dynamics radiates at kHz frequencies. The tidal disruption
that leads to the merger occurs above 600~Hz or so and the oscillations of the remnant could lead to a signal at
several kHz. However, the merger signal should be rich in information. In particular, it 
should tell us directly whether a massive neutron star or a black hole is formed, thus
placing constraints on the (hot) neutron star equation of state (EOS).
As we will discuss below, recent numerical relativity simulations  \cite{2003PhR...376...41B,BGR08,2009PhRvD..80f4037K,BGR09} illustrate the complexity of the merger
signal. On the one hand, this is problematic because we are unlikely to reach a stage where
we have truly reliable theoretical signal templates. On the other hand, 
one should be able to make progress by focussing on ``robust'' characteristics of the signal. 
However, it seems clear that we will need 3G detectors if we want to explore the 
science of the merger event. Combining the expected binary inspiral 
rates with the results in Figure \ref{fig:hs_psd_pol_vs_IF} we learn that, roughly, if the inspiral
phase is observable with Advanced LIGO/Virgo then ET should be able to detect the merger. 
This is strong motivation for not compromising on the high-frequency performance of a 3G detector.

The final stages of binary evolution before merger
may  also encode the detailed properties  of matter beyond nuclear
saturation \cite{2009PhRvD..79l4033R,2009JPhCS.189a2024M}. As the binary tightens the equation of state
begins to influence the GW signal, making it
possible to extract additional information (other than the masses and
spins). For mixed neutron-star/black-hole binaries, the characteristic frequency of tidal disruption, which
terminates the inspiral in some cases, also depends on the equation of
state  \cite{2006PhRvD..73b4012F,2008PhRvD..77h4002E,tidal1,tidal2,2009PhRvD..79d4030S}. The signal from a binary containing a strange quark star  would also be quite distinct in this 
respect \cite{2005PhRvD..71f4012L,2007PhR...442..109L,2008arXiv0801.4829G,2009arXiv0910.5169B}. For double neutron-star binaries, the tidal deformation due to
the other body can modify the orbital evolution in late inspiral. The
tidal phase corrections for most realistic equations of state, although invisible to
Advanced LIGO/Virgo, make potentially distinguishable contributions to the
signal in ET \cite{2009arXiv0911.3535H}.  The deformability of a given mass neutron star is encoded in the 
so-called Love numbers, which depend on the equation
of state and the resulting radius for a given mass \cite{2008ApJ...677.1216H,2008PhRvD..77b1502F}. Roughly, larger
neutron stars are deformed more easily. Further effects of matter dynamics, such as tidal
resonances, may also contribute to the signal \cite{1995MNRAS.275..301K,2006PhRvD..74b4007L,2007PhRvD..75d4001F}.

In order to illustrate the information that one may hope to
extract from  detecting the final
 merger of these sources we consider the GW 
emission produced by two equal-mass neutron stars. Detailed
simulations of such systems have been presented in~\cite{BGR08,BGR09}. Here we
simply recall that the simulations were performed using
high-resolution shock-capturing methods for the 
hydrodynamics equations and high-order finite-differencing techniques
for the Einstein equations. The complete set of
equations was solved using adaptive mesh-refinement techniques with
``moving boxes'' and the properties of the black holes produced in the
merger were extracted using the isolated-horizon formalism. The initial
data was obtained from a self-consistent solution of Einstein's equations in the
conformally-flat approximation and represents a system of binary
neutron stars in irrotational quasicircular orbits. The matter was
evolved using two idealized EOS given either by
the ``cold'' or polytropic EOS $p = K \rho^{\Gamma}$, or  the
``hot'' (ideal fluid) EOS $p = (\Gamma-1) \rho\, \epsilon$,
where $\rho$ is the rest-mass density, $\epsilon$ the specific
internal energy, $K$ the polytropic constant and $\Gamma$ the
adiabatic exponent. It should be noted that because the polytropic EOS
is isentropic it is unrealistic for describing the post-merger
evolution. Nevertheless, it provides a reasonably realistic description of
the inspiral phase, during which the neutron stars are expected to
interact only gravitationally. Moreover, because these two EOSs are
mathematically equivalent in the absence of shocks, one can 
 employ the \textit{same}
initial data and thus compare directly the influence that the EOS has
on the evolution. Also, because they represent
``extremes'' of the possible fluid behaviour, they offer insights into the expected range in dynamical behaviour.

In order to quantify the impact that different EOSs have on the
GW signal we  consider the power spectral density
(PSD) of the effective amplitude ${\tilde h}(f)$
\begin{equation}
{\tilde h}(f)\equiv \sqrt{\frac{{\tilde h}^2_{+}(f)+
	{\tilde h}^2_{\times}(f)}{2}}\,,
\end{equation}
where $f$ is the gravitational-wave frequency and where
\begin{equation}
{\tilde h}_{+,\times}(f) \equiv \int_0^{\infty}
	e^{2\pi i f t} h_{+,\times}(t) dt
\end{equation}
are the Fourier transforms of the gravitational-wave amplitudes
$h_{+,\times}(t)$, built using only the (dominant) $l=m=2$ multipole.

Fig.~\ref{fig:hs_psd_pol_vs_IF}  shows the PSD of the $l=m=2$
component of ${\tilde h}(f) f^{1/2}$ for a binary with total
gravitational mass $M=2.98\,M_{\odot}$ (``high-mass'' binary, left
panel) and for a binary with $M=2.69\,M_{\odot}$ (``low-mass''
binary, right panel) at a distance of  $300\,\mpc$. In both
cases the binaries are evolved from an initial separation of $45$~km.
To emphasize the differences induced by the EOS, we show the PSDs when the binary is evolved with either
the ``cold'' or  the ``hot'' EOS. For comparison, we also provide the result for an
equal-mass, nonspinning black-hole binary with total mass
$M=200\,M_{\odot}$ (also at a distance of $300\,\mpc$). Finally,  
we show the
phenomenological evolutionary tracks obtained by combining post-Newtonian and
numerical relativity results~\cite{Aetal07,Aetal08}. 

The PSD for the high-mass cold-EOS binary (blue curve in  the left panel
of Fig.~\ref{fig:hs_psd_pol_vs_IF}) is quite simple to interpret. It
shows, besides the large power at low frequencies corresponding to the
inspiral,  a peak at $f\approx 4\,\khz$ corresponding to the rapid
merger of the two neutron stars. For this cold EOS the merger
is not influenced by increased pressure forces via shock heating, so the
GW signal terminates abruptly with a prompt collapse
to black hole and a cut-off corresponding to the fundamental
quasi-normal mode (QNM) of the black hole at $f_{_{\rm QNM}} \simeq
6.7\,\khz$. In contrast, the PSD for the  high-mass hot-EOS binary 
is more complex, with the inspiral peak being
accompanied by a number of other peaks, the two most prominent, at $f\approx 1.75\,\khz$ and $f\approx
3\,\khz$, having
almost comparable amplitude. These additional peaks are related to the post-merger phase and the dynamics of the
hypermassive neutron star that is formed after the merger. The signal is sensitive to 
the dynamics of the cores of the two neutron stars,
which merge and ``bounce'' several times before the hypermassive neutron star eventually
collapses to a black hole, leaving a signature at $f\approx
4\,\khz$. The fundamental QNM frequency at $f_{_{\rm
    QNM}} \simeq 7.0\,\khz$ marks the cutoff of the signal also in this case.

We can interpret the PSDs of the low-mass
binaries in a similar way. The cold EOS results, in particular, show a very broad peak
between $f\approx 2\,\khz$ and $\approx 3.5\,\khz$  related to
the dynamics of the bar-shaped hypermassive neutron star that persists for several milliseconds after the merger. 
A small excess at $f\gtrsim 4\,\khz$  is associated with the
collapse to a black hole, whose fundamental QNM has a frequency of $f_{_{\rm QNM}} \simeq 7.3\,\khz$. 
Interestingly, the low-mass
hot EOS PSD does not show the broad peak. Instead, there is a  very narrow and
high-amplitude  peak around $f\approx 2\,\khz$. This feature is  related to a
 long-lived bar deformation of the hypermassive neutron star, which was evolved for
$\sim 16$ revolutions without the hypermassive neutron star collapsing to a black
hole. It should be noted that the simulations were performed assuming a
rotational symmetry~\cite{BGR08}, which prevents the growth
of the $m=1$ modes  which have been shown to limit the persistence of 
the bar-mode instability (see \cite{BdPMR07} for a detailed discussion). Although it is reasonable to expect that the bar
deformation will persist for several milliseconds after merger, it is
unclear whether this prominent peak will remain (and if so at what
amplitude) when the simulations are repeated with more generic
boundary conditions. The high-frequency part of
the PSD for the low-mass hot-EOS binary obviously does not show the
expected cut-off introduced by the collapse to black hole. An estimate
based on the secular increase of the central density suggests that, in this case the
collapse takes on a timescale of $\sim 110\,\ms$,
much longer than the simulations.

Three main conclusions can be drawn  from the results shown 
in Fig.~\ref{fig:hs_psd_pol_vs_IF}. First of all, with the
exception of very massive neutron stars (in which case the collapse of
the hypermassive neutron star  occurs essentially simultaneously with the merger), the
GW signal from binary neutron stars is considerably
richer (and more complex) than that from binary black holes. Secondly,
 while small differences between the two EOSs appear already
during the inspiral, it is really the post-merger phase that is
markedly different. Hence, an accurate description of the post-merger
evolution is \textit{essential} not only to detect this part of the
signal, but also to extract  information concerning the neutron star
interior structure. 
Finally, the parts of the PSD that
are most interesting and most likely to yield fundamental clues to the
physics beyond nuclear density, are likely to be only marginally detectable by 
 detectors like advanced LIGO/Virgo. As a result, 3G
detectors like ET may  provide the first realistic opportunity to use 
GWs as a Rosetta
stone to decipher the physics of neutron star interiors.

\section{Core collapse supernovae and hot remnants}

Neutron stars are born when a massive star runs out of nuclear fuel and collapses under
its own gravity. The dynamics of the stellar evolution after core bounce is tremendously complicated, and depends
on the interplay of a number of physical mechanisms. This complex process can produce GWs through
a number of different
channels, some of which are connected to the dynamics of the proto-neutron star and its immediate environment (usually associated
with high-frequency components of the signal), and others which depend on the convective zone behind the stalled
shock front (which gives rise to low frequency signals).
For slowly rotating iron cores, bounce and initial ringdown
are expected to lead to signals with peak frequencies in the range of $700 - 900 \, \mathrm{Hz}$ and dimensionless strain amplitudes 
of less than $5 \times 10^{-22}$ at a distance of 10~kpc \cite{2008PhRvD..78f4056D}, i.e. within our Galaxy. 
Faster rotation amplifies the bounce signal: If the iron core has moderate rotation, the peak frequencies span the 
larger range of $400 - 800 \, \mathrm{Hz}$ with amplitudes of $5 \times 10^{-22}$ up to $10^{-20}$. Very rapid
rotation leads to bounce at subnuclear densities, and GW  signals in a significantly lower frequency band of 
hundreds of Hz and strains around $5 \times 10^{-21}$ at 10~kpc.
% Anisotropic neutrino emission with frequencies of $20 - 200 \, \mathrm{Hz}$
% and Galactic strain amplitudes of $10^{-22}$ to $10^{-21}$ lie in a more sensitive band for detection, and for
% neutrino-driven core collapse supernovae these processes may be the primary source of detectable GWs. If  supernovae
% are actually driven by acoustic modes of the proto-neutron star, this part of signal may be competitive in terms of its signal-to-noise ratio, but it lies in a different frequency band. Finally, the accretion shock can be subject to another set of instabilities (standing accretion
% shock instabilities, SASI) which may emit roughly in the $100 - 800 \, \mathrm{Hz}$ band at up to $10^{-22}$ at $10 \, \mathrm{kpc}$.
Prompt convection occuring shortly after core bounce due to negative lepton gradients lead to Galactic signal amplitudes
in the range of $10^{-23}$ to $10^{-21}$ at frequencies of $50 - 1000 \, \mathrm{Hz}$ (all data is taken from
from table 2 in \cite{2009CQGra..26f3001O}, see also \cite{2006PhRvL..96t1102O, 2009A&A...496..475M}), 
whereas signals of convection in the proto-neutron star have strains of up to $5 \times 10^{-23}$ in a somewhat larger range
of frequencies. Neutrino-driven convection and a potential instability of the accretion-shock, the standing
accretion shock instability (SASI), could be relevant sources as well, with strain amplitudes up to $10^{-22}$ at
$100 - 800 \, \mathrm{Hz}$. In addition, an acoustic mechanism has been proposed for supernova
explosions \cite{2006ApJ...640..878B} which is connected with low-order g-mode oscillations in the
proto-neutron star. If this mechanism is active, very large strain amplitudes of up to $5 \times 10^{-20}$
at 10~kpc could be reached in extreme cases \cite{2006PhRvL..96t1102O}.

\begin{figure*}[t]
 \begin{center}
 \includegraphics[angle=-0,width=0.8\textwidth]{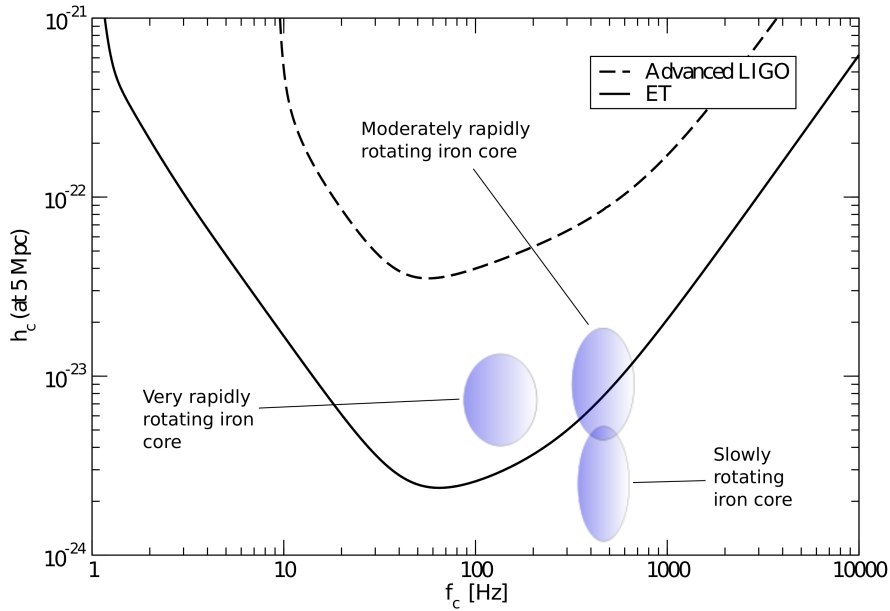}
 \end{center}
   \caption{Characteristic signal strengths, at a distance of 5~Mpc, associated with the bounce in core collapse, adapted
from fig. 3 in \cite{2009CQGra..26f3001O}. Typical iron cores have low rotation rates
and consequently weak core bounce signals. However, signals from systems with moderate or rapid rotation, which is expected
to be present in a subset of progenitor stars, can potentially be detected with ET
even at this distance.}
   \label{SNstrain}
\end{figure*}

A major uncertainty connected with supernova models is the initial state, in particular the angular momentum distribution
in the iron core. Current expectations from stellar evolution calculations imply a very slowly rotating core as a
canonical case \cite{2004ApJ...603..221M}. This is further supported by the observation that neutron stars seem to be born with comparatively low
rotation rates, and by recent evidence for massive loss of angular momentum
of stars before the white-dwarf stage \cite{2009Natur.461..501C}.  Strong GW signals can only be obtained by invoking processes which break the approximate
spherical symmetry of the system. If the core rotates faster than expected, for example for
collapsing stars that lead to gamma-ray bursts \cite{2006ApJ...637..914W}, then rotational
instabilities may become relevant sources of GWs. Of course,  these 
instabilities (or magnetic fields) must be effective enough to spin down young neutron stars after birth, in order to reconcile this scenario
with the observed neutron star spin distribution. Also, magnetic wind-up may in some cases
open channels to magnetically driven explosions, which could give rise to detectable signals.

The traditional bar-mode instability operates only at very high spin rates (measured in terms of the ratio of rotational
kinetic to gravitational binding energy, $T/|W|$, the limit is about $0.25$ in general relativity), which are not expected
even in more extreme models \cite{2008PhRvD..78f4056D}. Moreover, recent work \cite{2007PhRvD..75d4023B} suggests that once it is active
the instability does not persist for long due to nonlinear mode-mode coupling. Long-lived bar configurations require extreme 
fine-tuning of the initial data, something that nature is rather unlikely to provide. 
However, if the proto-neutron star is sufficiently differentially rotating,  so-called low $T/|W|$
instabilities may be operative \cite{2005ApJ...618L..37W}. This may lead to substantial deformations which could be detectable. The
associated signal-to-noise ratio depends (essentially) on 
the number of cycles and the saturation level of the instability. A typical value for the maximal strain
obtained in numerical simulations is $10^{-21}$ at $10 \, \mathrm{kpc}$
 with a frequency around $400 - 900 \, \mathrm{Hz}$ \cite{2007PhRvL..98z1101O}. 
Finally, secular instabilities (e.g. driven by the GW emission) could be active during or after the unbinding of the stellar envelope. We will discuss these instabilities later.

The different emission mechanisms all have quite characteristic signatures, so GW measurements
would provide
an unusually direct (probably the only besides neutrinos) way of probing the conditions inside core collapse supernovae \cite{2009CQGra..26f3001O}. However,  the
signal-to-noise ratio estimated
from numerical simulations make a detection of an extragalactic core collapse supernova from a slowly rotating (canonical) iron
core seem unlikely even with second generation detectors. Even if higher core rotation rates are assumed, detections
will be possible maybe up to at most 1 or $2 \, \mathrm{Mpc}$. Since the rates of (successful) supernovae are known from observations,
we know that Galactic events happen every $30 - 100$ years. Even at a distance of 1 Mpc, that is for very optimistic
GW estimates, the event rate is low enough that we
may have to be lucky to even
see a single event during the whole operation of Advanced LIGO. However, at a distance
of $3-5~\mathrm{Mpc}$, a range which could admit a detectable signal in a 3G detector, the event rate rate would be a few per year
\cite{2005PhRvL..95q1101A, 2008arXiv0810.1959K}. Therefore, it is safe to say that investigating
the core collapse supernova mechanism with GWs absolutely requires a 3G detector to obtain
meaningful statistics.

The proto-neutron star that is born in a core collapse is a
hot and rapidly evolving object.
After the first tenths of seconds of the remnant's life,  the lepton pressure
in the interior decreases
due to extensive neutrino losses,  the mantle  contracts and
the radius reduces to about $20-30$~km \cite{1986ApJ...307..178B}. This is known as a proto-neutron star.
The subsequent evolution  is ``quasi-stationary'', and can
be described by  a sequence of equilibrium configurations
\cite{pons1}.
Initially, the diffusion  of high-energy
neutrinos from the core to the surface
generates a large amount of heat within
the star, while the core entropy approximately doubles.
Within $\sim$~10~s the lepton content of the proto-neutron star is
drastically reduced. While the star is very hot,
neutrino pairs of all flavours are thermally produced
and dominate the emission. As
the neutrinos continue to diffuse and cool the star, their average
energy decreases and their
mean free path increases. After a few tens of seconds, the mean free
path becomes comparable
to the stellar radius, the net lepton number in the interior has
decreased to very low values,
the temperature has dropped to around $10^{10}$~K, and
the star becomes a neutrino-transparent neutron star.
The strong entropy and temperature gradients which develop in the interior
 during this evolution
strongly affect the star's oscillation frequencies.

As examples of this let us  consider the fundamental mode and the first gravity g-mode, cf. 
the discussion in \cite{hotstar1}. During the first 5~s of evolution the 
frequencies of these modes change dramatically. In the case of the f-mode, 
the frequency scales with the square  root of the average density so it will 
more than double as the star shrinks to the final 10~km radius. Meanwhile, the g-modes
depend on the strong temperature gradients that only prevail as long as the neutrinos are trapped. 
Detailed calculations \cite{hotstar1,hotstar2} confirm
that, during the first few tens of seconds,
the mode frequencies and damping times are considerably different
from those of a cold neutron star.  Moreover, dissipative processes
``competing'' with  GW emission,
(essentially  due to neutrino viscosity,
diffusivity, thermal conductivity, or thermodiffusion)
have timescales of the order
$t_{diss}  \approx 10-20$ seconds (see \cite{hotstar1} for details).
For instance, for the f-mode the GW damping time $t_f$ is such that
$t_f < t_{diss}$ when $t \gappreq 0.2$ seconds. 
Therefore, if some energy is initially stored into this mode,
it will be emitted in GWs. In contrast, the g-mode
is an efficient emitter of GWs only during the first
second, since after this time its damping time becomes
larger than $t_{diss}$. 

\begin{figure}[t]
\begin{center}
\includegraphics[width=0.6\textwidth]{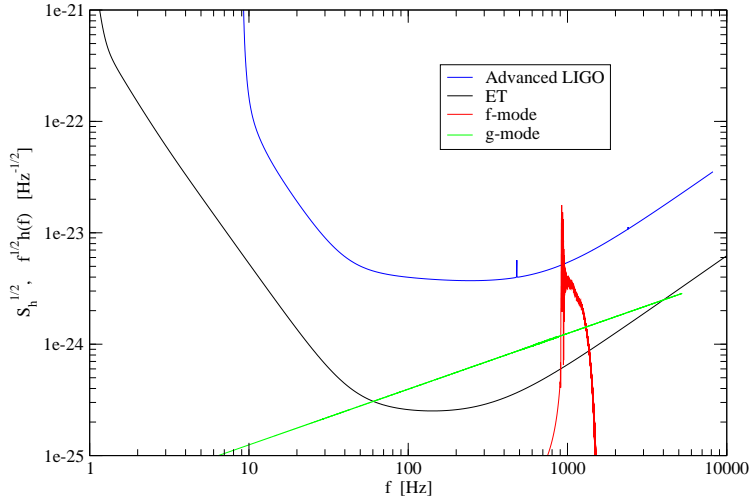}
\vspace{0.5cm}
\caption{ This figure compares the strain amplitude of the signal emitted by a 
proto-neutron star oscillating in either the f-mode
or  the g-mode to the sensitivity of Advanced LIGO and ET. 
The signal is assumed to have an amplitude at the detector site which
corresponds to an ET detection with a signal-to-noise ratio of 8 using matched filtering techniques.
 For a galactic source  at distance of 10~kpc, this would require an 
energy equivalent to $\Delta E_f \sim 2\times 10^{-12}~M_\odot c^2$ and 
$\Delta E_g \sim 3\times 10^{-11}~M_\odot c^2$ to be radiated through the 
f- and g-mode, respectively. A key point of these results is that the 
oscillation spectrum evolves during the observation. 
It is important to establish to what extent future detections, e.g. with ET, may be able to extract
the characteristics of these signals and allow us to probe the physics of  newly born neutron stars.
}
\label{figval}
\end{center}
\end{figure}

As we will discuss later, it is straightforward
to estimate the detectability of an oscillation mode once we know the associated 
frequency and damping time. Figure~\ref{figval} shows results from 
\cite{hotstar1} for the f- and the most promising g-mode of a proto-neutron star.
The data corresponds to signals detectable with a signal-to-noise ratio of 8 in ET. 
For a galactic source, at a distance of 10~kpc, this would require an 
energy equivalent to $\Delta E_f \sim 2\times 10^{-12}~M_\odot c^2$ and 
$\Delta E_g \sim 3\times 10^{-11}~M_\odot c^2$ to be radiated through the 
f- and g-mode, respectively (note that the signal strength scales with the square root of
the radiated energy). This level of energy releases is not unrealistic. 
Recent core collapse simulations suggest that
a conservative estimate of the  amount of energy emitted in 
GWs is of the order of $10^{-9}-10^{-8}~M_\odot c^2$. (As a comparison
and useful upper limit,
the collapse to a black hole of an old and rapidly rotating neutron star
would emit an energy in GWs of the order of $10^{-6}~M_\odot
c^2$~\cite{Baiotti06}.) 
If a  small fraction of this energy goes into the
excitation of the two modes shown in Figure~\ref{figval}, they may be observable.
Moreover, given such a signal one would expect to be able to identify
features in the waveform. 
For instance, one may infer the slope of the g-mode signal, or the detailed
structure of the f-mode waveform.
From these features one could hope to gain insight into  physical processes
occurring inside the star. In this way ET has the potential to be a
powerful instrument for exploring the physics of newly born neutron stars.
 
%%%%%%%%%%%%%%%%%%%%%%%%%%%%%%%%%%%%%%%%%%%%%%%%%%%%%%%%%%%%%%%%
%%%%%%%%%%%%%%%%%%%%%%%%%%%%%%%%%%%%%%%%%%%%%%%%%%%%%%%%%%%%%%%%

\section{Rotating deformed neutron stars}

Asymmetries, generated either by strains in the star's crust or by the
magnetic field, are expected to slowly leak rotational energy away
from spinning neutron stars. Such sources would be the
GW analogue of radio pulsars.  Indeed, the known radio
pulsar population, together with the accreting low-mass X-ray binary
systems, are prime candidates for GW detection via
targeted searches, where the observed electromagnetic phase is used to
guide the gravitational search.  Equally interestingly, there may be a
population of neutron stars currently invisible via electromagnetic
observations, spinning down by GW emission.  These
require an all-sky blind search, the computational costs of which are
very high, requiring the sort of computing power made available by the
{\tt Einstein@Home} project \cite{s4eah}.

\begin{figure}
  \epsfig{file=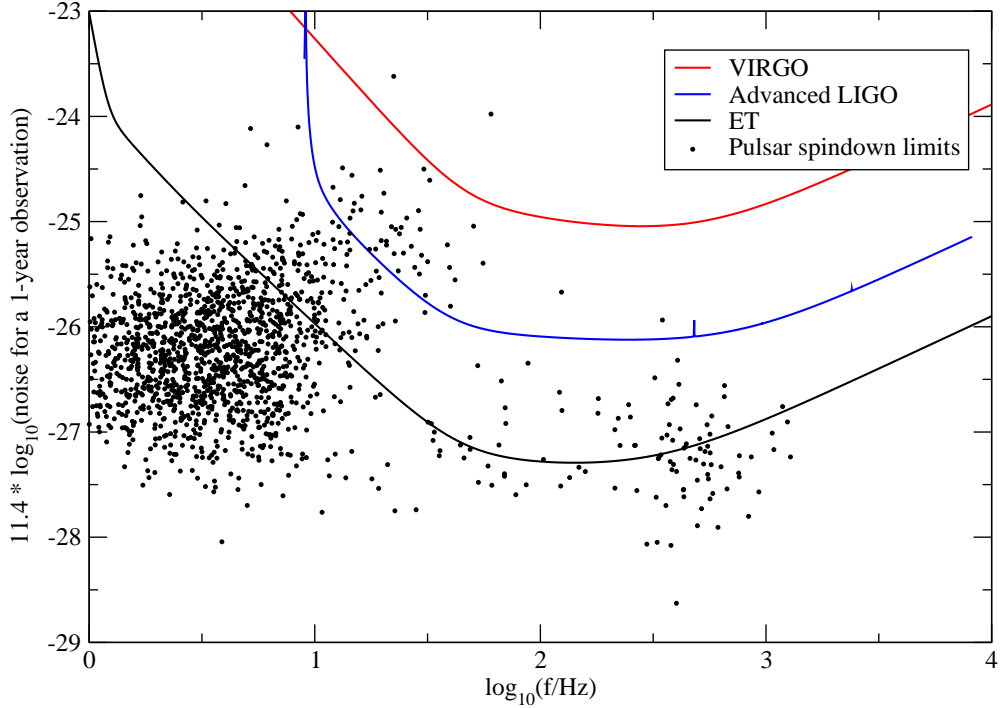,width=12cm,angle=-90}
\caption{Upper bounds on GW amplitude from known
  pulsars assuming 100\% conversion of spindown energy into GWs.
      An integration time of one year is assumed.
\label{fig:spindown}}
\end{figure}

How likely is a detection of a spinning deformed star, and how does ET
enter into the game?  For targeted searches for stars of known
distance, spin frequency and spin-down rate, one can place an upper
bound on the GW emission by assuming that all of the
kinetic energy being lost is being converted into GWs.  Such a plot is shown in Figure \ref{fig:spindown} for the
known pulsar population, with various detector noisecurves, including
ET, shown for reference.

Superficially, the figure would suggest that several tens of pulsars
might be detectable by Advanced LIGO/Virgo, with many more potentially
detectable by ET.  However, this plot makes a dangerous assumption: by
assuming 100\% conversion of kinetic energy into GW
energy, it implicitly requires neutron stars to be capable of
supporting asymmetries of sufficient size to power the necessary
GW emission.  The crucial question then becomes: what
level of asymmetry  would one expect a neutron star to have? This is a
complicated multi-faceted problem, where the answer depends not only
on the properties of the star, but also on the star's evolutionary
history.  So far, theoretical modelling has mainly focused on
establishing  what the largest possible neutron star ``mountain''
would be \cite{2000MNRAS.319..902U}.  Expressing this in terms of a (quadrupole) ellipticity, the
most detailed modelling of crustal strains suggest that \cite{2006MNRAS.373.1423H}
\begin{equation}
\epsilon < 2\times 10^{-5} \left( {u_\mathrm{break} \over 0.1} \right) ,
\label{maxeps}\end{equation}
where $u_\mathrm{break}$ is the crustal breaking strain.  Recent
molecular dynamics simulations \cite{2009PhRvL.102s1102H} suggest that this may be as large as
$0.1$, much larger than had been anticipated.  This would make the
neutron crust super-strong! In comparison, terrestrial materials have
$u_\mathrm{break} \approx 10^{-4} - 10^{-2}$. Basically, the available
estimates suggest that the crust would break if the deformation were
to exceed about 20~cm (on a 10~km star).

State-of-the-art calculations of the high density equation of
state suggest that  solid phases may also be present at higher
densities, allowing the construction of stars with larger
deformations.  Based on a model of a solid strange quark star, Owen \cite{2005PhRvL..95u1101O}
estimates $\epsilon < 6 \times 10^{-4} (u_\mathrm{break}/10^{-2})$,
while, based on a crystalline colour superconducting quark phase,
Haskell et al estimate $\epsilon < 10^{-3}
(u_\mathrm{break}/10^{-2})$ \cite{2007PhRvL..99w1101H}.  (Note that the molecular dynamics
simulations of \cite{2009PhRvL.102s1102H} do not apply to such exotic phases.)
So, significantly larger mountains \emph{might} be provided by nature,
depending upon the high density equation of state.

The magnetic field will also tend to deform the star, but for typical
pulsar field strengths the deformation is small \cite{2008MNRAS.385..531H,2008MNRAS.385.2080C,2009MNRAS.395.2162L}:
\begin{equation}
\epsilon \approx 10^{-12} \left(\frac{B}{10^{12} \rm \, G}\right)^2 \ .
\end{equation}
Note however that it is the \emph{internal}, rather than the external
field strength that counts.  Also, the above estimate assumes a normal
fluid core; a superconducting core complicates this picture, and could
produce larger asymmetries.  A simple estimate for a type II
superconducting core gives \cite{2002PhRvD..66h4025C,2008MNRAS.383.1551A}
\begin{equation}
\epsilon \approx 10^{-9}
\left(\frac{B}{10^{12} \rm \, G}\right)
\left(\frac{H_{\rm crit}}{10^{15} \rm \, G}\right) \ ,
\end{equation}
where $H_{\rm crit}$ is the so-called critical field strength \cite{2002PhRvD..66h4025C}.

\begin{figure}
  \epsfig{file=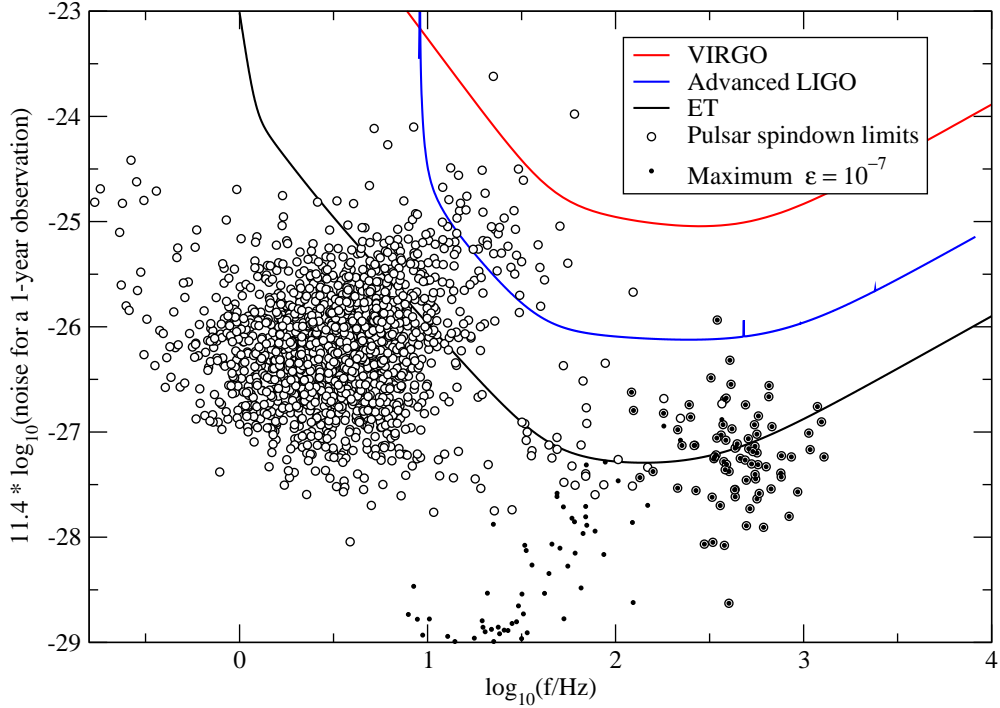,width=12cm,angle=-90}
\caption{Spindown upper limits on known radio pulsars.  The open circles
    assume 100\% coversion of spindown energy into
    GWs, regardless of how large an ellipticity is
    required to power such emission.  In contrast, the open circles
    limit the allowed ellipticity to be no greater than $10^{-7}$.  An
    integration time of one year is assumed.
 \label{fig:max_epsilon}}
\end{figure}

Clearly, this is a rather complicated story, with different physical
assumptions leading to very different possible maximum mountain sizes.
To gain some understanding of how the  maximum mountain size
affects detection prospects, in Figure \ref{fig:max_epsilon} we re-plot
the spin-down limits of Figure \ref{fig:spindown}, this time limiting
the maximum mountain size to $10^{-7}$, a possibly optimistic but
certainly not unrealistic value.  We plot the original spin down
limits of Figure \ref{fig:spindown} as open circles, and the limits
obtained by putting a $10^{-7}$ cut-off in ellipticity as solid
circles.

Clearly, Figure \ref{fig:max_epsilon} tells a rather different story
from Figure \ref{fig:spindown}.  Most of the younger pulsars, on the
left of the diagram, have dropped off completely: the ellipticites
needed for them to radiate all of their kinetic energy via
GWs are unphysically large.  In contrast, the
millisecond pulsars on the right hand side have stayed put: they are
able to spin down via the GW channel with ellipticites
smaller than $10^{-7}$.  Interestingly, many  of the millisecond pulsars lie below the
Advanced LIGO/Virgo noise curves but above the ET one.  So, if nature
supplies millisecond pulsars deformed at the level of one part in
$10^7$, ET may provide the key to detecting them.

This dependence on rather uncertain physics makes the prospect of a
detection all the more exciting.  In fact, observations of targeted
radio pulsars are already providing interesting results. An
observational milestone was reached recently, when LIGO  data from the first 9 months of the S5 science run was used to beat the Crab pulsar spin-down limit \cite{ligo_crab}.  Since then a larger data set, taken from the full S5 run, has been analysed;  it was found that no more than
$2\%$ of the spin-down energy was being emitted in the GW channel, corresponding to an ellipticity bound of approximately $\epsilon < 10^{-4}$ \cite{2009arXiv0909.3583T}.  This result shows that GW emission
does not dominate the spin-down of the Crab.  As argued in \cite{2000A&A...354..163P}, there was no
real possibility that 100\% of the spin-down was GW
powered, as pure gravitational spin-down would conflict with the Crab's measured braking index.  However, the fact that the GW contribution to spin-down  is less than $2\%$ was not at all 
obvious. It tells us, for instance,  that the Crab is not a maximally
strained quark star.

How are these results likely to improve in the future? This is quite
easy to estimate since the sensitivity of a search increases in
inverse proportion to the detector noise level and as the square root
of the observation time.  This means that, in the case of the Crab, a
search over two years of Advanced LIGO/Virgo data would be sensitive to
mountains of about $10^{-5}$, while ET may push the limit to
$10^{-6}$, a sufficiently small value that even a conventional neutron
star crust may radiate at a detectable level. Similarly, ET may be able to 
detect deformations at the $\epsilon \sim 10^{-9}$ level in some of the millisecond pulsars.
One would probably expect a signal to be detected before this level is
reached, but unfortunately we do not know this for sure. The main
challenge concerns the generation mechanism for deformations.  Why is
the neutron star deformed in the first place? This is an urgent
problem that needs to be addressed by theorists.  A key question
concerns the size of the ``smallest'' allowed mountain. Based on current
understanding, the interior magnetic field sets the lower limit, as
given above.  Clearly, for a typical pulsar this limit is $\epsilon
\approx 10^{-12}$, too small to ever be detected.  Of equal interest
is the likely -- as opposed to maximum -- size of an elastic mountain, but
this is an even more difficult problem, depending not just upon the
material properties of the crust but also on its ``geological'' history.

As far as evolutionary scenarios are concerned, accreting neutron
stars in low-mass X-ray binaries have attracted the most attention. 
This is natural for a number of reasons.  First of all, the currently
observed spin distribution in these systems seems consistent with the
presence of a mechanism that halts the spin-up due to accretion well
before the neutron star reaches the break-up limit  \cite{1998ApJ...501L..89B}.
GW emission could provide a balancing torque if the accretion leads
to non-axisymmetries building up in the crust
\cite{1998ApJ...501L..89B, 2000MNRAS.319..902U,2009MNRAS.395.1972V,2005ApJ...623.1044M,2009arXiv0910.5064W}; the required
deformation is certainly smaller than (\ref{maxeps}).  Alternatively,
accreting stars might rotate fast enough for some modes of oscillation
to go unstable, again provide a gravitational spin-down torque to
oppose the accretion spin-up \cite{1999ApJ...516..307A,1999ApJ...517..328L,2002MNRAS.337.1224A}.  The problem is that accreting
systems are very messy.  In particular, we do not understand the detailed
accretion torque very well \cite{2005MNRAS.361.1153A}. To make progress we need to improve our
theoretical models, and we also need future high precision X-ray
timing observations to help constrain the binary parameters. 

%---

Unlike say binary neutron star mergers or stellar collapse, the
detection of long-lasting periodic GWs from a deformed
neutron star would be a significant data analysis and computational
problem even if we had perfect theoretical templates.  This warrants
an extended discussion of the data analysis problems that must be
addressed before data from Advanced LIGO/Virgo and ET  can be fully exploited
for continuous wave signal detection and astrophysics.  The difficulty arises
because of a combination of the expected low signal-to-noise ratio and the very large parameter space of possible signal
shapes.  Here we briefly review existing search techniques and discuss
the improvements necessary in the ET era, to detect continuous wave signals.

Ignore the various systematic signal model uncertainties for the
moment, and consider searching for the signals using different
techniques, ranging from optimal matched filtering to time-frequency
methods.  The GW phase model typically considered for these searches
is, in the rest frame of the neutron star
\begin{equation}
  \label{eq:phasemodel}
  \Phi(\tau) = \Phi_0 + 2\pi \left[f(\tau-\tau_0) +
    \frac{\dot{f}}{2}(\tau-\tau_0)^2 + \ldots  \right]\,.
\end{equation}
Here $\tau$ is time in the rest frame of the star, $\Phi_0$ the
initial phase, $\tau_0$ a fiducial reference time, and $f$, $\dot{f}$
are respectively the instantaneous frequency and its time derivative
at $\tau_0$.  Going from $\tau$ to detector time $t$ requires us to
take into account the motion of the detector in the solar system and,
if the neutron star is in a binary system, the parameters of its
orbit.  

Using  matched filtering, the signal-to-noise ratio builds up
with the observation time $T$ as $T^{1/2}$.  More precisely, the
minimum detectable amplitude $h_0^{min}$ for a single template search
at a frequency $f$ for $D$ detectors, each with a single sided noise
power spectral density $S_n(f)$ is
\begin{equation}
  \label{eq:h0min}
  h_0^{min} = 11.4 \sqrt{\frac{S_n(f)}{DT}}\,.
\end{equation}
Here we have chosen thresholds corresponding to a false alarm rate of
$1\%$ and a false dismissal rate of $10\%$, and averaged over all
possible pulsar orientations and sky positions.  For the isolated
neutron stars discussed previously, if a fraction $p$ of their
spindown energy goes into GWs, the required observation time for a
detection is proportional to $S_n(f)/(Dp)$.  Thus, if we wish to probe
the Crab pulsar at say the $p=1\%$ level as opposed to the $6\%$ limit
set by the LIGO S5 search, we would need to
observe six times longer.  This longer observation time leads to much
more stringent requirements on the accuracy of the signal model used in
the search.  Presently, one assumes that the GWs are locked to the EM
phase. Do we really believe that this assumption holds over say, a 5 year
observation period, corresponding to a frequency offset of $\sim 6
\times 10^{-9}\,$Hz?  Furthermore, there might be interesting physics
causing the GW frequency to be slightly different from twice the radio
signal frequency.  The accuracy with which we can determine this
deviation (and the parameters of any model describing these
deviations) is proportional to the inverse of the signal-to-noise
ratio (SNR). We could, even without a 3G detector, try to improve the
SNR by increasing the observation time.  However, a larger observation
time typically implies a larger parameter space to search. This makes
the search computationally harder and also adversely affects the
statistical significance of any discoveries.  This argues in favour of
increasing the SNR by either increasing the number of detectors or
making them more sensitive.  The same holds, in fact to a much greater
degree, for the wide parameter space searches as we now discuss.

If we do not know the signal parameters \emph{a priori}, then we need
to perform a search over a parameter space informed by astrophysical
constraints.  The most straightforward way of doing the search is to
lay a grid of templates over this parameter space and to find the
template leading to the largest value of the appropriate detection
statistic.  It turns out that in most cases, the number of templates
increases very quickly with $T$. This search over a large number of
templates has two main effects, both of which affect the sensitivity
adversely.  The first is that the larger number of statistical trials
leads to a correspondingly larger probability of a statistical false
alarm or equivalently, to a larger threshold if we want to maintain a
fixed false alarm rate.  The second and more important effect is
simply that the number of templates might become so large that the
computational cost becomes a key factor  limiting the largest
value of $T$ that we can consider.  For isolated neutron stars, the
relevant parameters in a blind search are the sky position and the
frequency and higher time derivatives of the frequency.  For a pulsar
in a binary system, the orbital parameters need to be considered as
well, leading to three additional parameters for circular orbits (five if we
include eccentricity).  A detailed analysis, including a catalogue of
accreting neutron stars of potential interest for GW
searches can be found in \cite{lmxbdetect}.

This data analysis challenge needs to be addressed with better
techniques and larger computational resources.  Examples of more
efficient data analysis techniques are the so-called semi-coherent
methods which break up the data set of duration $T$ into $N$ smaller
segments of length $T_{coh}$ and combine the results of a coherent
matched filter search from each segment.  While such a method would
be suboptimal when we are not computationally limited, it would lead
to a closer to optimal method in the presence of computational
constraints. Examples of such methods are described in
\cite{stackslide, hough, virgohough}, and
examples of GW searches employing them are \cite{s2hough, s4psh, s4eah, s5powerflux}.  
The sensitivity of these searches is
\begin{equation}
  \label{eq:h0min-semicoh}
  h_0^{min} = \frac{k}{N^{1/4}}\sqrt{\frac{S_n(f)}{DT_{coh}}}
\end{equation}
where $k$ is typically $\sim 30$. The goal here is to make $T_{coh}$
as large as computational resources allow.  Further developments of
these techniques (a notable example is \cite{pletschallen}), and
the ability to follow up candidate events enabling multi-stage hierarchical searches (see e.g. \cite{hierarchical}) will be necessary for both
isolated and binary systems.  A different kind of
technique is based on cross-correlating data from multiple detectors.
These are again less sensitive than coherent matched filtering, but
they are computationally easy and highly robust against signal
uncertainties.  A description of an algorithm tailored towards continuous wave
signals can be found in \cite{crosscorr}.  It is shown that this
method is, in a suitably generalized sense, also an example of a
semi-coherent technique.  This approach is expected to be especially valuable for
binary systems where there are potential uncertainties in the signal
model. It has, in fact, already been applied to LIGO S4 data
\cite{s4radiometer}.

The continued development of larger computational platforms, and
increasing the available computational resources faster than the
baseline improvement predicted by Moore's law will be critical.  A
good example of a large  platform tailored towards such
searches is the \texttt{Einstein@Home} project, and the increased use
of GPUs is another. On the timescale relevant for ET we expect to see significant
improvement of our computational resources.

One crucial point is that there is more to the detection process than
simply observing the amplitude of the signal -- higher precision will
allow you to extract more physics.  Even in the simple case of a
steadily rotating star, getting an accurate fix on the phase allows
detailed comparison between the electromagnetic
and gravitational signals, something which can provide information (in
a model dependent way) on the coupling between the stellar components
and possibly on the nature of pulsar timing noise \cite{2004PhRvD..70d2002J}.  Alternatively, if
the rotating star undergoes free precession or contains a pinned
superfluid component misaligned with the principal axes, there will be
radiation at multiple harmonics, some of which might be significantly
weaker than others \cite{2002MNRAS.331..203J, 2005CQGra..22.1825V}.  Observing these harmonics, and comparing their
sizes, will yield yet more information on the stellar interior \cite{2009arXiv0909.4035J}.  
Higher quality observations can also be used to measure the distances to (sufficiently nearby) pulsars \cite{2005PhRvD..71l3002S}, thus allowing an estimate of the ellipticity to be made using
GW data alone and helping us to build a three dimensional picture of the source distribution.  The
key point for instruments like ET is this: it is not enough to simply detect a
signal -- the higher the signal-to-noise, the more detailed a picture
of the waveform  can be built up.  This provides strong motivation
for a 3G search for GWs, even if second
generation detectors prove successful in identifying signals.

\section{Oscillations and instabilities}

In principle, the most promising strategy for constraining the physics of neutron stars  involves
observing their various modes of oscillation. We have already discussed the particular case of newly born neutron stars. The problem for mature neutron
stars is similar, in the sense that different
families of modes can be (more or less) directly associated with different core physics. For example, the
fundamental f-mode (which should be the most efficient GW emitter) scales with average density,
while the pressure p-modes overtones probe the sound speed throughout the star, the gravity g-modes are sensitive to thermal/composition gradients
and the w-modes represent oscillations of spacetime iself.
A mature neutron star also has elastic shear modes in the crust \cite{2007MNRAS.374..256S} and superfluid modes associated with
regions where various constituents form large scale condensates \cite{2008PhRvD..78h3008L}. Magnetic stars may have complex dynamics
due to the internal magnetic field.
Finally, in a rotating star, there is a large class of inertial modes which are restored by the Coriolis force. 
One of these rotationally restored modes is the r-mode. The r-mode is particularly interesting because it
may be driven unstable by the emission of gravitational radiation.

While the asteroseismology strategy seems promising \cite{1996PhRvL..77.4134A,1998MNRAS.299.1059A,2001MNRAS.320..307K,2004PhRvD..70l4015B,2007GReGr..39.1323B,2008GReGr..40..945F} (provided we 
can model neutron stars at a sufficient level of realism), it is clear that 
it relies on how well future GW detectors will be able to
detect the various pulsation modes. It is quite easy to make this 
question quantitative since a typical GW signal from a
neutron star pulsation mode will take the
form of a damped sinusoid, with oscillation frequency $f$ and damping time $t_d$. That is, we would have
\begin{equation}
h(t) = {\cal A} e^{-(t-T)/t_d} \sin [ 2\pi f (t-t_0)] \quad \mbox{ for }
       t > t_0
\end{equation}
where $t_0$ is the arrival time of the signal at the detector (and 
$h(t)=0$ for $t<t_0$).
Using standard results for the GW flux 
the amplitude ${\cal A}$ of the signal can be expressed in terms of the
total energy radiated in the oscillation,
\begin{equation}
{\cal A} \approx 7.6\times 10^{-24} \sqrt{{\Delta E_\odot \over
10^{-12} } {1 \mbox{ s} \over t_d}}
 \left( {1 \mbox{ kpc} \over d } \right)
\left({ 1 \mbox{ kHz} \over f} \right)  \ .
\end{equation}
where $\Delta E_\odot = \Delta E/M_\odot c^2$.
Finally, the signal-to-noise ratio for this signal can be
estimated from
\begin{equation}
\left({S \over N} \right)^2 = { 4Q^2 \over 1+4Q^2} {{ \cal A}^2 t_d
\over 2S_n}
\label{sign}\end{equation}
where the ``quality factor'' is $Q=\pi f t_d$ and
$S_n$ is the spectral noise density of the detector. It is worth noting that, for
oscillations lasting longer than the observation time $T$ (i.e. when $Q\gg1$)
we regain the continuous-wave result that the signal-to-noise ratio improves as the square root of $T$. 

So far, the key parameters  are the oscillation frequency and 
the damping time of the mode. We now need an astrophysical scenario that 
excites the oscillations. Moreover, this scenario has to be such that it can be 
modeled using linear perturbation theory (with a  ``slowly'' evolving
background configuration as reference). This means that  one would not expect to 
deal with the violent dynamics immediately following the formation of a hot neutron star, 
either after binary merger or core collapse. As we have already 
discussed, these problems are the realm of nonlinear simulations.
Once the (proto-)neutron star settles down to a relatively slow evolution, 
the mode-problem becomes relevant. Of course, in this regime it is less obvious
that the oscillations will be excited to large amplitude. Hence, 
the interest in the various instabilities that may affect the star as it 
evolves is natural. We can also ask general questions concerning 
the amount of energy that would need to be channeled
through the various modes, and whether these energy levels are
astrophysically ``reasonable''. An optimistic scenario would perhaps
consider that 
as much as $10^{-6}M_\odot c^2$ may be radiated (this would be the energy level expected from collapse to a black hole \cite{Baiotti06}).
However, this number is likely
only relevant (if at all) for the oscillations of a hot remnant. 
For mature neutron stars, significantly lower energy levels should be 
expected. We can take as a bench-mark the energy involved in a typical 
pulsar glitch, in which case one might (still optimistically?) radiate
an energy equivalent to $10^{-13}M_\odot c^2$ \cite{2001PhRvL..87x1101A}.

\begin{figure}
  \epsfig{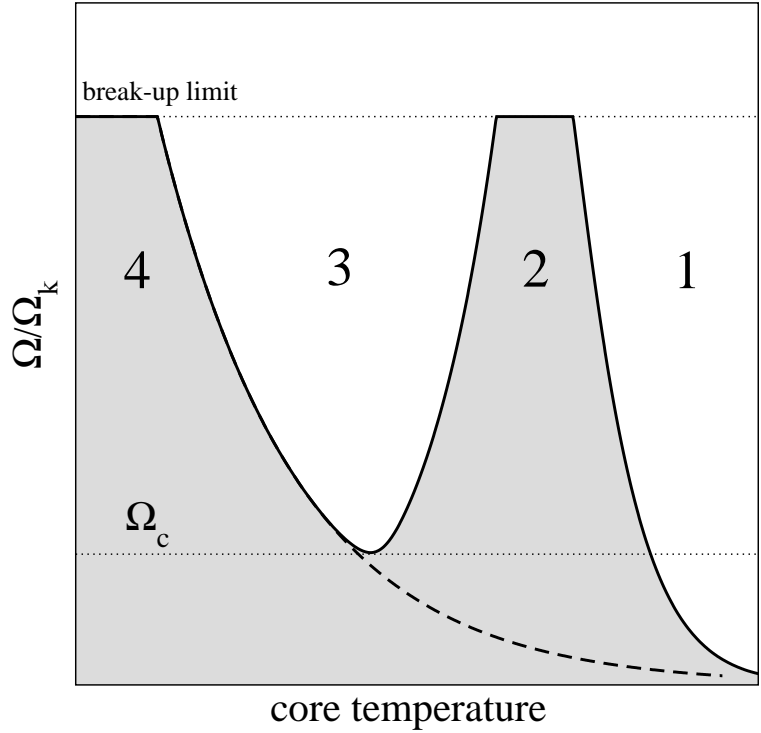}
\caption{This figure provides a schematic illustration of the r-mode instability window.
At low temperatures 
(region 4) dissipation due to shear viscosity (dashed curve) counteracts the instability.
At temperatures of the order of $10^{10}$~K bulk viscosity suppresses
the instability (region 2). The associated critical temperature is due to a ``resonance'' 
between the oscillation and  nuclear reaction timescales. 
At very high temperatures (region 1) the nuclear reactions that lead to 
the bulk viscosity are suppressed and an unstable mode can, in principle, grow. 
However, this region may only be relevant 
for the first few tens of seconds following the birth of a neutron star.
The main instability window is expected at temperatures near 
$10^9$~K (region 3), above a critical rotation $\Omega_\mathrm{c}$.  Provided that gravitational radiation 
drives the unstable mode strongly enough the instability should 
govern the spin-evolution of a neutron star. 
The instability window may change considerably
if we  add more detailed
pieces of physics, like superfluidity and the 
presence of hyperons, to the model. }
  \label{instab}
\end{figure}

As far as instabilities are concerned, the GW driven instability
of the r-mode remains (after more than a decade of scrutiny) the most promising \cite{1998ApJ...502..708A,1998ApJ...502..714F,1998PhRvL..80.4843L,1999ApJ...510..846A,2001IJMPD..10..381A,cqg}.
The r-mode instability window depends on a balance between GW driving and various dissipation mechanisms.
In principle, this provides a sensitive probe of the core physics. 
To illustrate this let us consider a simple model of a neutron star composed
of neutrons, protons and electrons, ignoring issues to do with the crust physics, 
superfluidity, magnetic fields etcetera. If we take the overall density profile to be that of an 
$n=1$ polytrope (a simple yet useful approximation) then the characteristic
growth timescale for the 
$l=m=2$ r-mode is
\begin{equation}
t_{\rm gw} \approx  50  \left( {1.4 M_\odot \over M } \right) \left( {10\ \mathrm{km} \over R} \right)^4 \left( {P \over 1\ \mathrm{ms} } \right)^6 \ \mbox{s} 
\end{equation}
where $P$ is the spin-period of the star.
In the simplest model the unstable mode is damped by  shear and bulk viscosity.
At relatively low temperatures (below a few times $10^9$~K) 
the main viscous dissipation mechanism  arises from
momentum transport due to particle scattering,  modelled as a
macroscopic shear viscosity. In a normal fluid star neutron-neutron 
scattering 
provides the most important contribution. This leads to a 
typical damping time
\begin{equation}
t_{\rm sv} \approx 7\times10^7 \left( {1.4 M_\odot \over M } \right)^{5/4} \left( { R \over 10\ \mathrm{km}} \right)^{23/4} \left({ T \over 10^9\ \mathrm{K} } \right)^2 \mbox{ s}
\label{sv1}\end{equation}
In other words, at a core temperature of $10^9$~K the damping timescale is 
longer than a year.
At higher temperatures
bulk viscosity is the dominant dissipation
mechanism. Bulk viscosity arises as
the mode oscillation
drives the fluid away from beta equilibrium.
It depends on the extent to which energy is
dissipated from the fluid motion as weak interactions
try to re-establish equilibrium. This is essentially a resonant mechanism that 
is efficient when the oscillation timescale is similar to the reaction timescale. 
At higher and lower frequencies (or, equivalently, temperatures) the bulk viscosity mechanism 
shuts off. This resonance is apparent in the schematic instability window shown in Figure~\ref{instab}.
In the Cowling approximation one can show that the bulk viscosity 
damping timescale (relevant just above $10^9$~K, i.e. in region 3 in  Figure~\ref{instab})
is given by
\begin{equation}
t_{\rm bv} \approx 3 \times10^{11} \left( { M \over 1.4 M_\odot} \right) \left({ 10\ \mathrm{km} \over R} \right)\left( {P \over 1\ \mathrm{ms} } \right)^2 \left({ 10^9\ \mathrm{K} \over T} \right)^6
\mbox{ s}
\label{bulkest}\end{equation}
It is easy to see that, while this damping is inefficient at $10^9$~K it is very efficient at $10^{10}$~K.

From these estimates we learn that the r-mode
instability will only be active in a certain temperature range.
To have an  instability we need
$t_{\rm gw}$ to be  smaller in magnitude than both $t_{\rm sv}$ and 
$t_{\rm bv}$. We find that 
 shear viscosity will completely suppress the 
r-mode instability at core temperatures below $10^5$~K.
This corresponds to region 4 in Figure~\ref{instab}. 
Similarly, bulk viscosity will prevent the mode from growing 
in a star that is hotter than a few times
$10^{10}$~K. This is the case in region 2 of Figure~\ref{instab}.
However, if the core becomes very hot (as in region 1 in Figure~\ref{instab}), 
then the nuclear reactions that lead to the bulk 
viscosity are strongly suppressed. For our chosen model, this region is likely not very relevant 
since neutron stars will not remain
at such extreme temperatures long enough that an unstable mode can grow 
to a large amplitude. Having said that, one should be aware that the ``resonance temperature'' where the 
bulk viscosity is strong is highly model dependent and there may well be situations where region 1 
is relevant. Finally,  in the intermediate region
3 in figure~\ref{instab}, there is a temperature window  where the growth time due
to gravitational radiation is short enough to overcome the
viscous damping and  the mode is unstable. 

This schematic picture holds  for all modes that are driven unstable by GWs, but the 
 instability windows are obviously different in each case.  
Detailed calculations show that 
viscosity stabilizes all f-modes below $\Omega_c \approx 0.95\Omega_K$
in a Newtonian star, while the r-modes are stable below $\Omega_c \approx 0.04\Omega_K$. (Note that the instability windows
are likely to be significantly smaller in more realistic models.)
For the $n=1$ polytrope that we have considered the break-up limit $\Omega_K$  would correspond to a 
spin period of around 1.2~ms (for uniform rotation).
Because of the potentially very large instability window,  the r-modes have been studied
in a variety of contexts,  at many different levels of detail. (It is perhaps worth noting that, with the exception of \cite{2001PhRvD..63b4019L,2001MNRAS.328..678R,2002MNRAS.330.1027R,2003PhRvD..68l4010L}, most of the
relevant studies have not accounted for relativistic gravity.) We now (think we) know that key issues concern the interaction with magnetic
fields in the star~\cite{Rezzolla00,Rezzolla01a,Rezzolla01b},
the damping due to the vortex mediated mutual friction in a superfluid \cite{1991ApJ...380..530M,2000PhRvD..61j4003L,2006MNRAS.368..162A,2009MNRAS.397.1464H}, the role of turbulence \cite{2009arXiv0911.1609M}, the boundary layer at the crust-core interface \cite{2000PhRvD..62h4030L,2006PhRvD..74d4040G,2006MNRAS.371.1311G,2006ApJ...644L..53P} and
exotic bulk viscosity due to the presence of hyperons \cite{2002PhRvD..65f3006L,2006PhRvD..73h4001N} or deconfined quarks \cite{1998PhRvL..81.3311M,2000PhRvL..85...10M} in the deep neutron star core. These
problems are all very challenging. In addition, we need to  model the GW signal from an unstable r-mode.
This is also difficult because, even though the r-mode growth phase is adequately described by linear theory, nonlinear
effects soon become important \cite{1998PhRvD..58h4020O}. Detailed studies show that the instability saturates at a low amplitude due to coupling to other inertial modes 
\cite{2002ApJ...571..435M,2002PhRvD..65b4001S,2003ApJ...591.1129A,2004PhRvD..70l1501B,2004PhRvD..70l4017B,2005PhRvD..71f4029B}. 
The subsequent evolution is very complex, as is the associated GW signal \cite{2002CQGra..19.1247O,2007PhRvD..76f4019B,2009PhRvD..79j4003B}.

While we improve our understanding of this mechanism, we should not forget about the f-mode instability. 
Basically, we know that the strongest f-mode instability in a Newtonian model is associated with the $l=m=4$ modes \cite{1991ApJ...373..213I}. 
The situation is different in a relativistic model where the $l=m=2$ modes may also be unstable \cite{1998ApJ...492..301S}. The quadrupole modes
are more efficient GW emitters and so could lead to an interesting instability. Moreover, the supposed ``killer blow''
of superfluid mutual friction suppression \cite{1995ApJ...444..804L} only acts below the critical temperature for superfluidity. 
This means that, even in the worst case scenario, the f-mode could experience an instability in very young 
neutrons stars (provided that they are born spinning fast enough) \cite{2009PhRvD..79j3009A}. This problem clearly needs 
further attention. In particular, it should be noted that the various dissipative mechanisms have only been discussed in 
Newtonian gravity. This means that we have no realistic  estimates of the damping of the unstable quadrupole mode. 
Moreover, not much effort has gone into understanding the associated saturation amplitude and 
potential GW signal.

To model a truly realistic oscillating neutron star may be difficult, but the potential reward is considerable.
This is  clear from recent results for the quasiperiodic oscillations that have been observed in the tails
of magnetar flares. These oscillations have been interpreted as torsional oscillations of the crust.
These crust modes should be easier to excite than poloidal
oscillations since they do not involve density variations.
Three giant flare events, with peak luminosities in the range 10$^{44}$ - 10$^{46}$ erg s$^{-1}$,
have been observed so far; SGR 0526-66 in 1979, SGR 1900+14 in 1998, and SGR 1806-20 in 2004.
Timing analysis of the last two events revealed several quasiperiodic oscillations
in the decaying tail, with frequencies in the expected range for 
toroidal crust modes \cite{Israel2005,SW2005,WS2006a,SW2006,WS2006b}.
If this interpretation is correct, then we are already doing neutron star asteroseismology!
Subsequent calculations have shown how
observations may constrain both the mass, radius and crust thickness of these stars \cite{SA2007,SA2009}. 
Eventually, it may also be possible to probe the magnetic field configuration \cite{2008MNRAS.385.2161S}. There are, however, 
important conceptual issues that we need to resolve in order to make progress on this problem. A key question 
concerns the  existence of a magnetic continuum \cite{2006MNRAS.368L..35L,2007MNRAS.377..159L, 2007MNRAS.375..261S,2007MNRAS.374.1015L,2008MNRAS.385.2069L,2008MNRAS.385L...5S,2009MNRAS.395.1163S,2009MNRAS.396.1441C,2009MNRAS.397.1607C} and its potential effect on the dynamics of the system.

Our current models may not be particularly reliable,
but they should motivate us to improve our understanding of the key physics (like the interior magnetic field \cite{Bfield1}, superfluid phases \cite{2008MNRAS.391..283V,2009MNRAS.396..894A} and
the dynamical coupling between the crust and the core \cite{2001PhRvD..64d4009M}). It is also possible that the magnetar events, such as the 27/12/2004 burst in SGR 1806-20,
generate GWs. Indeed, LIGO data has already been searched for signals from soft-gamma ray repeaters \cite{Abbott2007,Abbott2008,2009ApJ...701L..68A}, including searches at frequencies at which f-modes might be expected to radiate.  No detections have yet been made.
Is this surprising? Given our current uncertainty as to just which parts of the star are set in motion  following a burst, probably not. Basically, pure crust oscillations would not generate strong GWs 
due to the low density involved. The situation may change if the dense core fluid is involved in the
oscillation, e.g. through magnetic coupling \cite{2006MNRAS.368L..35L,GSA2006},
but in that case we do not yet know what the exact signature of the event would be so it is difficult
to search for. It would most likely not be appropriate to search for a signal at the observed frequencies in the X-ray data. 
We need to make progress on a number of issues if we want to understand these events.
In absence of better information perhaps the best we can do is ``look under the light''.

\section{Challenges}

GW astronomy promises to provide insights into the ``dark side'' of the Universe.
Because of their density, neutron stars are ideal GW
sources and we hope to be able to probe the extreme physics of their interior with future detectors.
The potential for this is clear, in particular with third generation detectors like the Einstein Telescope.
However, in order to detect the signals and extract as much information as possible, we need to improve our
theoretical models considerably.
It seems natural to conclude this survey of scenarios by discussing some of 
the main challenges that we need to meet in the next few years.

For binary inspirals, we need to work out when finite
size effects begin to affect the evolution. We should consider tidal resonances and compressibility
in detail and ask at what level they affect the late stages of inspiral. It is also important to 
quantify to what extent we will be able to ``read off'' neutron star physics from
a detected signal. 

For hot young remnants,
resulting from binary mergers or core collapse events, we need to refine our large scale numerical simulations.
It is important to understand that, despite great progress in recent years, these simulations are not yet in the truly ``convergent'' regime.
Key technical issues (like how one should
model, and keep track of, the fluid to vacuum transition at the surface) still need to be resolved. Future
simulations must use ``realistic'' equations of state, and consider evolving composition,
heat/neutrino cooling and magnetic fields with as few ``cheats'' as possible. To some extent this effort may be helped
by the fact that many effects that are important in a mature neutron are less so on the dynamical timescales 
that lend themselves to large-scale simulations. 

For rotating deformed neutron stars, we need to go beyond producing
upper limits on possible signal strengths, and address the much more
subtle issue of their \emph{likely} deformations.  It will also be vitial
for  the source modelling community to engage further with the high
energy physics/QCD theorists, whose models of matter at high density
could prove crucial in interpreting future detections.  Finally, the
data analysis challenge involved in looking for the long-lived signals
expected from rotating stars is formidable --- further work on reducing
the  computational burden of these searches is crucial.

In parallel, we need to
build on our current understanding of neutron star oscillations and instabilities. This effort should aim
at accounting for as much of the interior physics as possible. If we want to model proto-neutron stars then we 
need to refine our understanding of heat conduction (in general relativity) and the role of the trapped neutrinos. 
For mature neutron stars we need to model the crust, which will contain a ``decoupled'' neutron superfluid. 
The complex composition of the core, with potentially several exotic phases of matter, poses a number of challenges. 
To meet these we need to improve our models for multi-fluid systems. It is also key to understand the relevance
of proton superconductivity. After all, the magnetic field will always play some role. For magnetars it may, in fact, 
be the dominant factor.  Finally, it is crucial that we find a way to model systems that evolve on a secular timescale. 
The archetypal problem may be the nonlinear saturation and subsequent evolution of an unstable r-mode. At the present time, 
the available models are based on the explicit calculation of, and transfer of energy between, coupled modes. 
It will not be easy to extend this calculation to 
more realistic neutron star models, and it seems natural to ask if there is some alternative 
way of modelling such problems.

Finally, we need a clearer
phenomenological understanding of pulsar glitches, accreting neutron stars, magnetar flares etcetera.
It is natural to develop this as part of an effort in multi-messenger
astronomy \cite{2008cosp...37.2316O,2009CQGra..26t4014O}, a theme we
have only touched upon briefly.  Electromagnetic and
GW data have in fact already been usefully combined.  For
instance, LIGO data was used to show that the gamma ray burst
GRB070201, which originated in the direction of the spiral arms of the
Andromeda galaxy, was \emph{not} powered by a neutron star inspiral at the distance of Andromeda \cite{2008ApJ...681.1419A}.
The improved sensitivity of 3G detectors will allow for
more frequent and more powerful statements to be made, covering a
whole host of electromagnetic observations.  In particular, it seems appropriate to ask what we can learn from more detailed radio observations 
with SKA and Lofar (for example). Future X-ray data should also continue to provide key information --- just think how much we have learned
from RXTE. In fact, accurate neutron star timing data is key for GW science. 
Of course, we should not simply wait for our colleagues in other fields to give us answers. We should  
ask if our modelling effort can help shed light on phenomena in ``mainstream'' astronomy. Given the accurate neutron star models
that we are now developing, it would seem obvious that this is the case.

\acknowledgements
NA and DIJ acknowledges support from STFC in the UK via grant number PP/E001025/1.
KDK and BZ are supported by the German Science Foundation (DFG) via 
SFB/TR7. We would like to thank Thomas Janka, Andrew Melatos, Ben Owen and B. Sathyaprakash for helpful comments. 
We also acknowledge many fruitful discussions with our colleagues in the Compstar network.

\end{document}